\documentclass[prd,aps,a4paper,superscriptaddress,twocolumn,nofootinbib]{revtex4}
\usepackage{graphicx}
\usepackage{color}
\usepackage{dcolumn}
\usepackage{bm}
\usepackage{slashed}
\usepackage{amsmath}
\usepackage{latexsym}
\usepackage{amssymb}
\usepackage{mathrsfs}
\usepackage{amsfonts}
\usepackage{url}
\usepackage{longtable}
\allowdisplaybreaks
\begin{document}
\title{On the transverse-traceless gauge condition when matters are presented}

\author{Yadong Xue} 
\affiliation{Institute for Frontiers in Astronomy and Astrophysics, Beijing Normal University, Beijing 102206, China}
\affiliation{Department of Astronomy, Beijing Normal University, Beijing 100875, China}
\author{Xiaokai He} 
\affiliation{School of Mathematics and Statistics, Hunan First Normal University, Changsha 410205, China}
\author{Zhoujian Cao
\footnote{corresponding author}} \email[Zhoujian Cao: ]{zjcao@amt.ac.cn}
\affiliation{Institute for Frontiers in Astronomy and Astrophysics, Beijing Normal University, Beijing 102206, China}
\affiliation{Department of Astronomy, Beijing Normal University, Beijing 100875, China}
\affiliation{School of Fundamental Physics and Mathematical Sciences, Hangzhou Institute for Advanced Study, UCAS, Hangzhou 310024, China}

\begin{abstract}
The transverse-traceless gauge condition is an important concept in the theory of gravitational wave. It is well known that vacuum is one of the key conditions to guarantee the existence of the transverse-traceless gauge. Although it is thin, interstellar medium is ubiquitous in the universe. Therefore, it is important to understand the concept of gravitational wave when matter is presented. Bondi-Metzner-Sachs theory has solved the gauge problem related to gravitational wave. But it does not help with the cases when gravitational wave propagates in matters. This paper discusses possible extensions of the transverse-traceless gauge condition to Minkowski perturbation with matter presented.
\end{abstract}

\maketitle

\section{Introduction}

Gauge issue is a long-standing problem in general relativity. Related to gravitational wave (GW), gauge issue ever confused people including Einstein for almost half a century \cite{kennefick2007traveling}. For isolated gravitational wave sources which correspond to asymptotically flat spacetime, Bondi-Metzner-Sachs theory \cite{BonVanMet62,Sac62} solves the gauge problem elegantly and gives a beautiful description of gravitational wave. In the cases of asymptotically non-flat spacetimes, including the cosmological situations, metric perturbation with special gauge choice is used to describe gravitational wave \cite{2005NJPh....7..204F,2022GReGr..54...89C}.

Since asymptotically flat spacetime can be approximated as a Minkowski perturbation, the transverse-traceless (TT) gauge condition becomes a convenient tool to describe gravitational wave. Many textbooks use the TT gauge condition as the basic theory to present gravitational wave \cite{liangcanbin,wald84,MTW73}. The TT gauge condition provides a fundamental picture of gravitational wave to many researchers.

It is well known that the TT gauge condition is a special harmonic gauge. It is the vacuum Einstein equation that aids us to pick out the special one from harmonic gauges. Although many text books use plane gravitational waves as examples to illustrate the choice procedure of the TT gauge condition, plane wave is not a necessity which we will clarify this issue later in the current paper. In contrast, vacuum condition must be satisfied to guarantee the existence of the TT gauge condition. Our analysis later will make this fact apparent.

In real situations, we need to consider that gravitational wave propagates in matters \cite{PhysRevD.94.024048,2022CQGra..39g5014L}. This is because that interstellar medium is ubiquitous in the universe although it is thin. And we need to consider the interaction between gravitational waves and matters including the Weber bar gravitational wave detectors \cite{maggiore2008gravitational}, GW-excited lunar vibrations \cite{Li:2023plm,PhysRevD.110.043009} and others. In the existing literature, people neglect the matter completely when they describe the gravitational waves in these situations. Intuitively we can believe this treatment is reasonable and quantitatively accurate enough. But it is still interesting to ask the mathematical principle behind. In fact Prof. Yuan-Zhong Zhang ever discussed this problem with one of us several times.

In the current paper we aim to discuss the concept of gravitational wave when matter is presented. More specifically we will discuss the generalization of the TT gauge condition to situations when matter is presented. In the next two sections we describe the mathematics needed by our analysis. Although the mathematical principle has been scattered in the existing literature, the description provided here is more systematic and comprehensible. Based on this mathematical tool, we briefly review the decomposition of the metric perturbation of Minkowsky in Sec.~\ref{sec4}. Besides the introduction of the well known gauge invariant quantities, we design three gauge sensitive variables which are useful to discuss the gauge conditions later. After that we give a clearer analysis of harmonic gauge conditions from a new viewpoint. Within this framework we discuss the TT gauge condition choice in Sec.~\ref{sec6}. Along with that a natural extension of the usual TT gauge condition to situations when matters are present is proposed. Alternatively we analyze the CZ gauge condition proposed by other authors in Sec.~\ref{sec7}. Intuitively people may think the CZ gauge condition is similar but different to harmonic gauge condition. Our analysis indicates that the CZ gauge is more similar to the usual TT gauge condition. The similarity is that harmonic gauge is a family of gauges while the usual TT gauge and the CZ gauge are unique. Our analysis shows that the CZ gauge in general is not harmonic. But the CZ gauge condition returns to TT gauge condition when the spacetime is vacuum. In this sense the CZ gauge condition can also be looked as a generalization of TT gauge condition to situations when matters are present. We summarize the paper with some discussions about the generalized TT gauge condition at last section.

The geometric units with $c=G=1$ are used throughout the paper.

\section{decomposition of a vector field}

For any vector field $\vec{v}$ with asymptotic condition $\vec{v}(\vec{r}\rightarrow\infty)=0$ we have following decomposition
\begin{align}
\vec{v}&=\nabla \phi+\nabla\times\vec{u},\label{eq2}
\end{align}
where the scalar field $\phi$ is determined by
\begin{align}
\nabla^2\phi&=\nabla\cdot\vec{v}.\label{eq1}
\end{align}
According to the partial differential equation theory for Poisson equation, there is an unique solution for $\phi$ to the above equation. As the solution of the above mentioned Poisson equation, asymptotically we have $\phi(\vec{r}\rightarrow\infty)=0$.

Then the decomposition condition Eq.~(\ref{eq2}) leads us to
\begin{align}
\nabla\times\vec{u}&=\vec{v}-\nabla \phi,\\
\nabla\times(\nabla\times\vec{u})&=\nabla\times\vec{v}-\nabla\times(\nabla \phi)=\nabla\times\vec{v},\\
\nabla(\nabla\cdot\vec{u})-\nabla^2\vec{u}&=\nabla\times\vec{v}.
\end{align}
We assume $\nabla\cdot\vec{u}=0$, thus the last equation becomes
\begin{align}
\nabla^2\vec{u}&=-\nabla\times\vec{v}.\label{eq3}
\end{align}
According to the partial differential equation theory for Poisson equation again, there is an unique solution for $\vec{u}$ to the above equation. And asymptotically we have $\vec{u}(\vec{r}\rightarrow\infty)=0$.

Based on Eq.~(\ref{eq3}) we have
\begin{align}
\nabla\cdot\nabla^2\vec{u}&=-\nabla\cdot(\nabla\times\vec{v})=0\\
\nabla^2(\nabla\cdot\vec{u})&=0.
\end{align}
Using the partial differential equation theory for Laplace equation, we confirm
\begin{align}
\nabla\cdot\vec{u}=0,
\end{align}
which satisfies the requirement to get Eq.~(\ref{eq3}).

In another words, any vector $\vec{v}$ with asymptotic condition $\vec{v}(\vec{r}\rightarrow\infty)=0$ can be uniquely decomposed to a transverse part $\nabla\times\vec{u}$, where $\vec{u}$ is determined by Eq.~(\ref{eq3}) satisfying $\nabla\cdot\vec{u}=0$, and longitudinal part $\nabla \phi$, where $\phi$ is determined by Eq.~(\ref{eq1}) .

The decomposition Eq.~(\ref{eq2}) admits an interesting property which is related to harmonic gauge condition involved in the gravitational wave theory. In order to illustrate this property we explain a theorem first. Considering a Poisson equation
\begin{align}
\nabla^2u=s,
\end{align}
if the source term $s$ is a harmonic function which means $\Box s=0$, then
\begin{align}
&\Box\nabla^2u=\Box s=0,\\
&\nabla^2\Box u=0.
\end{align}
Based on asymptotic boundary condition, the above equation means $\Box u=0$. This is to say the solution of a Poisson equation with a harmonic source is a harmonic function.

The above mentioned theorem together with Eqs.~(\ref{eq1}) and (\ref{eq3}) tell us that if the original vector field $\vec{v}$ is harmonic $\Box\vec{v}=0$, then the decomposition components $\phi$ and $\vec{u}$ are also harmonic satisfying $\Box\phi=0$ and $\Box\vec{u}=0$.

Actually the aforementioned property can even be stronger as following. Assuming that $\phi$ and $\vec{u}$ are decomposition components of a given field $\vec{v}$, then $\Box\phi$ and $\Box\vec{u}$ are decomposition components of the vector field $\Box\vec{v}$.

\section{decomposition of a tensor field}\label{sec3}

Closely following the trick of the above decomposition of a vector field, for any tensor field $h_{ij}$ with asymptotic condition $h_{ij}(\vec{r}\rightarrow\infty)=0$ we have following decomposition.
\begin{align}
h_{ij}&=\frac{1}{3}H\delta_{ij}+\left[\partial_i\partial_j-\frac{1}{3}\delta_{ij}\nabla^2\right]\lambda+\partial_i\epsilon_{j}+\partial_j\epsilon_{i}+h_{ij}^{\rm TT}\label{eq4}
\end{align}
where the scalar fields $H$ and $\lambda$ are determined by
\begin{align}
H&=\delta^{ij}h_{ij},\label{eq6}\\
\nabla^2\tau&=\frac{3}{2}\partial^j\partial^i\left(h_{ij}-\frac{1}{3}H\delta_{ij}\right)=\frac{3}{2}\partial^j\partial^ih_{ij}-\frac{1}{2}\nabla^2H,\label{eq7}\\
\nabla^2\lambda&=\tau\label{eq8}.
\end{align}
According to the partial differential equation theory for Poisson equation, there are unique solution for $\tau$ and $\lambda$ to the above equations. And asymptotically we have $H(\vec{r}\rightarrow\infty)=0$, $\tau(\vec{r}\rightarrow\infty)=0$ and $\lambda(\vec{r}\rightarrow\infty)=0$.

The transverse traceless notation TT in Eq.~(\ref{eq4}) means
\begin{align}
\partial^ih_{ij}^{\rm TT}&=0,\\
\delta^{ij}h_{ij}^{\rm TT}&=0.
\end{align}
Consequently Eq.~(\ref{eq4}) leads us to
\begin{align}
&\partial_i\epsilon_{j}+\partial_j\epsilon_{i}=h_{ij}-\frac{1}{3}H\delta_{ij}-\left[\partial_i\partial_j-\frac{1}{3}\delta_{ij}\nabla^2\right]\lambda-h_{ij}^{\rm TT},\\
&\nabla^2\epsilon_{j}+\partial_j\partial^i\epsilon_{i}=\partial^ih_{ij}-\frac{1}{3}\partial_jH-\left[\nabla^2\partial_j-\frac{1}{3}\partial_j\nabla^2\right]\lambda,\\
&\nabla^2\epsilon_{j}+\partial_j\partial^i\epsilon_{i}=\partial^ih_{ij}-\frac{1}{3}\partial_jH-\frac{2}{3}\partial_j\tau.\label{eq19}
\end{align}
In addition we require $\epsilon_i$ is divergence free
\begin{align}
\partial^i\epsilon_{i}=0.
\end{align}
Consequently Eq.~(\ref{eq19}) becomes
\begin{align}
\nabla^2\epsilon_{j}&=\partial^ih_{ij}-\frac{1}{3}\partial_jH-\frac{2}{3}\partial_j\tau.\label{eq5}
\end{align}
According to the partial differential equation theory for Poisson equation, there is an unique solution for $\epsilon_j$ to the above equation. At the mean time Eq.~(\ref{eq5}) leads to
\begin{align}
\partial^i\nabla^2\epsilon_i&=\partial^j\partial^ih_{ij}-\frac{1}{3}\nabla^2H-\frac{2}{3}\nabla^2\tau,\\
\nabla^2\partial^i\epsilon_i&=0.
\end{align}
Using the partial differential equation theory for Laplace equation, we know solution $\epsilon_i$ is automatically divergence free satisfying the assumption we made before. This equivalently means the tensor $\partial_i\epsilon_{j}$ is traceless.

Based on the determined $H$, $\lambda$ and $\epsilon_i$, we can determine $h^{\rm TT}_{ij}$ as
\begin{align}
h_{ij}^{\rm TT}\equiv h_{ij}-\frac{1}{3}H\delta_{ij}-\left[\partial_i\partial_j-\frac{1}{3}\delta_{ij}\nabla^2\right]\lambda-2\partial_{\left(i\right.}\epsilon_{\left.j\right)}.\label{eq9}
\end{align}
When $h_{ij}$ is symmetric, $h^{\rm TT}_{ij}$ is symmetric, transverse and traceless \cite{2005NJPh....7..204F,2009arXiv0912.0128R,PhysRevD.83.061501}.

Specifically for a planar gravitational wave
\begin{align}
h_{ij}=A_{ij}e^{I\vec{n}\cdot\vec{x}},
\end{align}
thus Eq.~(\ref{eq6}) gives us
\begin{align}
&H=Ae^{I\vec{n}\cdot\vec{x}},\\
&A\equiv\delta^{ij}A_{ij}.
\end{align}
Here we have used $I$ to denote $\sqrt{-1}$. Plugging the above relations into Eq.~(\ref{eq7}) we get
\begin{align}
&\nabla^2\tau=-\frac{3}{2}n^in^jh_{ij}+\frac{1}{2}n^2H,\\
&\tau=\frac{3}{2n^2}n^in^jh_{ij}-\frac{1}{2}H.
\end{align}
Continually plugging the above relations into Eq.~(\ref{eq8}) we get
\begin{align}
&\lambda=-\frac{3}{2n^4}n^in^jh_{ij}+\frac{1}{2n^2}H.
\end{align}
Plugging the above relations into Eq.~(\ref{eq5}) we get
\begin{align}
&\nabla^2\epsilon_j=In^ih_{ij}-In_j\frac{1}{n^2}n^kn^lh_{kl},\\
&\epsilon_j=-I\frac{n^i}{n^2}h_{ij}+I\frac{n_j}{n^4}n^kn^lh_{kl}.
\end{align}
Collecting these results into Eq.~(\ref{eq9}) we get
\begin{align}
&h_{ij}^{\rm TT}\equiv h_{ij}-n_in^kh_{kj}-n_jn^lh_{il}\nonumber\\
&-\frac{1}{2}\left(H\delta_{ij}-Hn_in_j-\delta_{ij}n^kn^lh_{kl}-n_in_jn^kn^lh_{kl}\right).
\end{align}

This is to say $h^{\rm TT}_{ij}$ can also be alternatively expressed as \cite{maggiore2008gravitational}
\begin{align}
&h_{ij}^{\rm TT}=\Lambda_{ij}{}^{kl}h_{kl},\\
&\Lambda_{ij}{}^{kl}\equiv P_i{}^kP_j{}^l-\frac{1}{2}P_{ij}P^{kl},
\end{align}
where the projector operator $P_i{}^k$ is defined as
\begin{align}
&P_i{}^k\tilde{h}_{kj}(\vec{n})\equiv\tilde{h}_{ij}(\vec{n})-n_in^k\tilde{h}_{kj}(\vec{n})
\end{align}
respect to the Fourier component $\tilde{h}_{kj}(\vec{n})$
\begin{align}
&h_{ij}\equiv\int\tilde{h}_{ij}(\vec{n})e^{I\vec{n}\cdot\vec{x}}d^3\vec{n}.
\end{align}

\section{decomposition of four dimensional rank 2 tensor}\label{sec4}

Given a Minkowski perturbation $h_{\mu\nu}$ satisfying $h_{\mu\nu}(\vec{r}\rightarrow\infty)\rightarrow0$, we can decompose its components with the techniques shown above as following \cite{2005NJPh....7..204F}
\begin{align}
&h_{tt}=2\phi,\label{eq22}\\
&h_{ti}=\beta_i+\partial_i\gamma,\\
&h_{ij}=h^{\rm TT}_{ij}+\frac{1}{3}H\delta_{ij}+\partial_{\left(i\right.}\epsilon_{\left.j\right)}+\left(\partial_i\partial_j-\frac{1}{3}\delta_{ij}\nabla^2\right)\lambda.\label{eq23}
\end{align}
Different to Sec.~\ref{sec3} we have absorbed the factor 2 into $\epsilon_i$ here.

Similarly we have decomposition for infinitely small gauge transformation $\xi_\mu$ satisfying $\xi_{\mu}(\vec{r}\rightarrow\infty)\rightarrow0$
\begin{align}
&\xi_t=A,\\
&\xi_i=B_i+\partial_iC.
\end{align}

Similar to the property of vector decomposition shown in the previous section, if $h_{\mu\nu}$ is harmonic, then all the decomposition components are harmonic. If the infinitely small gauge transformation $\xi_\mu$ is harmonic, the components $A$, $B_i$ and $C$ are also harmonic.

Under such gauge transformation $\xi_\mu$, the aforementioned decomposition components of $h_{\mu\nu}$ will change as
\begin{align}
&\phi\rightarrow\phi-\dot{A},\\
&\beta_i\rightarrow\beta_i-\dot{B}_i,\\
&\gamma\rightarrow\gamma-A-\dot{C},\\
&H\rightarrow H-2\nabla^2C,\\
&\lambda\rightarrow\lambda-2C,\\
&\epsilon_i\rightarrow\epsilon_i-2B_i,\\
&h^{\rm TT}_{ij}\rightarrow h^{\rm TT}_{ij}.
\end{align}
Based on the above transformation rules for the decomposition components, we can easily construct gauge invariant quantities \cite{2005NJPh....7..204F}
\begin{align}
&\Phi\equiv-\phi+\dot{\gamma}-\frac{1}{2}\ddot{\lambda},\\
&\Theta\equiv\frac{1}{3}\left(H-\nabla^2\lambda\right),\\
&\Xi_i\equiv\beta_i-\frac{1}{2}\dot{\epsilon}_i,\label{eq26}\\
&h^{\rm TT}_{ij}.
\end{align}
Straightforward calculation shows that the linearized Einstein tensor can be expressed with the above gauge invariant quantities as \cite{2005NJPh....7..204F}
\begin{align}
&G_{tt}=-\nabla^2\Theta,\\
&G_{ti}=-\frac{1}{2}\nabla^2\Xi_i-\partial_i\dot{\Theta},\\
&G_{ij}=-\frac{1}{2}\Box h^{\rm TT}_{ij}+\delta_{ij}\left[\frac{2}{3}\nabla^2(\Phi+\frac{1}{2}\Theta)-\ddot{\Theta}\right]
-\partial_{\left(i\right.}\dot{\Xi}_{\left.j\right)}\nonumber\\
&-\left(\partial_i\partial_j-\frac{1}{3}\delta_{ij}\nabla^2\right)(\Phi+\frac{1}{2}\Theta).
\end{align}
It can be shown more that the above form is also the decomposition form of the Einstein tensor.

We decompose the stress-energy tensor $T_{\mu\nu}$ also as
\begin{align}
&T_{tt}=\rho,\\
&T_{ti}=S_i+\partial_iS,\\
&T_{ij}=T^{\rm TT}_{ij}+P\delta_{ij}+\partial_{\left(i\right.}\sigma_{\left.j\right)}+\left(\partial_i\partial_j-\frac{1}{3}\delta_{ij}\nabla^2\right)\sigma.
\end{align}
One interesting point is that $T_{\mu\nu}$ is gauge invariant up to high order approximation. This is because $T_{\mu\nu}$ is a first order small quantity in our cases and the gauge transformation matrix of infinitely small gauge transformation is a sum of a identity matrix and a first order small quantity.

Based on the decomposition form of $G_{\mu\nu}$ and $T_{\mu\nu}$, the Einstein equation $G_{\mu\nu}=8\pi T_{\mu\nu}$ can be expressed as
\begin{align}
&\nabla^2\Theta=-8\pi\rho,\label{eq10}\\
&\nabla^2\Xi_i=-16\pi S_i,\label{eq11}\\
&\dot\Theta=-8\pi S,\label{eq12}\\
&\Box h^{\rm TT}_{ij}=-16\pi T^{\rm TT}_{ij},\\
&\frac{2}{3}\nabla^2(\Phi+\frac{1}{2}\Theta)-\ddot{\Theta}=8\pi P,\label{eq13}\\
&\dot{\Xi}_i=-8\pi\sigma_i,\label{eq28}\\
&\Phi+\frac{1}{2}\Theta=-8\pi\sigma.
\end{align}
Combining Eqs.~(\ref{eq10}), (\ref{eq12}) and (\ref{eq13}) we get
\begin{align}
\nabla^2\Phi=4\pi(\rho+3P-3\dot{S}).\label{eq16}
\end{align}

Regarding $T_{\mu\nu}$ the conservation law $\nabla^\mu T_{\mu\nu}=0$ leads us to \cite{2005NJPh....7..204F}
\begin{align}
&\dot{\rho}-\nabla^2S=0,\label{eq20}\\
&\frac{3}{2}P-\frac{3}{2}\dot{S}+\nabla^2\sigma=0,\\
&2\dot{S}_i-\nabla^2\sigma_i=0.\label{eq21}
\end{align}
These three relations guide us to introduce three gauge sensitive variables
\begin{align}
\Pi&\equiv\nabla^2\gamma-\dot{\phi}-\frac{1}{2}\dot{H},\\
\Lambda&\equiv-\frac{1}{4}H+\frac{3}{2}\phi+\nabla^2\lambda-\frac{3}{2}\dot{\gamma},\\
\Sigma_i&\equiv\dot{\beta}_i-\frac{1}{2}\nabla^2\epsilon_i\\
&=\dot{\Xi}_i+\frac{1}{2}\ddot{\epsilon}_i-\frac{1}{2}\nabla^2\epsilon_i\\
&=\dot{\Xi}_i-\frac{1}{2}\Box\epsilon_i\label{eq30}\\
&=-8\pi\sigma_i-\frac{1}{2}\Box\epsilon_i.
\end{align}
The second line of the above equation about $\Sigma_i$ is due to Eq.~(\ref{eq26}). The last line of the above equation about $\Sigma_i$ is due to Eq.~(\ref{eq28}). Under the gauge transformation defined by $\xi_\mu$, these gauge sensitive variables will change as
\begin{align}
&\Pi\rightarrow\Pi-\Box A,\\
&\Lambda\rightarrow\Lambda-\frac{3}{2}\Box C,\\
&\Sigma_i\rightarrow\Sigma_i+\Box B_i.
\end{align}

Given $\Xi_i$ and $\Sigma_i$ we can construct $\epsilon_i$ and $\beta_i$ as following. Based on Eq.~(\ref{eq30}) we have
\begin{align}
\Box\epsilon_i=2(\dot{\Xi}_i-\Sigma_i).
\end{align}
Up to initial data of $\epsilon_i$ and $\dot{\epsilon}_i$ we can get $\epsilon_i$. Then based on Eq.~(\ref{eq26}) we have
\begin{align}
\beta_i=\Xi_i+\frac{1}{2}\dot{\epsilon}_i.
\end{align}

Given $\Theta$, $\Phi$ and $\Lambda$ we can construct $\lambda$ through
\begin{align}
\Box\lambda=\Theta+2\Phi+\frac{4}{3}\Lambda.
\end{align}
Then $\lambda$ is determined up to initial data of $\lambda$ and $\dot{\lambda}$. Based on the determined $\lambda$ and the given $\Theta$ we can determine $H$ as
\begin{align}
H=3\Theta+\nabla^2\lambda.
\end{align}
Based on the determined $\lambda$ and $H$ and the given $\Pi$ and $\Phi$ we can construct $\gamma$ through
\begin{align}
\Box\gamma=\Pi-\dot{\Phi}+\frac{1}{2}\dot{H}-\frac{1}{2}\dddot{\lambda}.
\end{align}
Then $\gamma$ is determined up to initial data of $\gamma$ and $\dot{\gamma}$. Finally we can determine $\phi$ through
\begin{align}
\phi=\dot{\gamma}-\frac{1}{2}\ddot{\lambda}-\Phi.
\end{align}
\section{harmonic gauge conditions}

The linearized Einstein equation under the harmonic gauge condition reads as
\begin{align}
\Box \bar{h}_{\mu\nu}=-16\pi T_{\mu\nu},
\end{align}
here $\bar{h}_{\mu\nu}\equiv h_{\mu\nu}-\frac{1}{2}h\eta_{\mu\nu}$ means the corresponding trace-reversed metric perturbation of $h_{\mu\nu}$. The relation between the four dimensional trace $h$ and three dimensional trace $H$ is $h=H-2\phi$. Corresponding to Eqs.~(\ref{eq22})-(\ref{eq23}) we have
\begin{align}
&\bar{h}_{tt}=\phi+\frac{1}{2}H,\\
&\bar{h}_{ti}=\beta_i+\partial_i\gamma,\\
&\bar{h}_{ij}=h^{\rm TT}_{ij}+\frac{1}{3}(3\phi-\frac{1}{2}H)\delta_{ij}+\partial_{\left(i\right.}\epsilon_{\left.j\right)}\nonumber\\
&+\left(\partial_i\partial_j-\frac{1}{3}\delta_{ij}\nabla^2\right)\lambda.
\end{align}
The harmonic gauge condition $\eta^{\mu\sigma}\partial_\mu\bar{h}_{\sigma\nu}=0$ can be expressed as
\begin{align}
&\nabla^2\gamma-\dot{\phi}-\frac{1}{2}\dot{H}=0,\label{eq24}\\
&\frac{2}{3}\nabla^2\partial_i\lambda+\frac{1}{2}\nabla^2\epsilon_i+\partial_i\phi-\frac{1}{6}\partial_iH-\dot{\beta}_i-\partial_i\dot{\gamma}=0.\label{eq25}
\end{align}

The linearized Einstein equation under harmonic gauge condition becomes
\begin{align}
&\Box \phi=-4\pi(\rho+3P),\label{eq14}\\
&\Box\beta_i=-16\pi S_i,\\
&\Box \gamma=-16\pi S,\\
&\Box H=24\pi(P-\rho),\\
&\Box\lambda=-16\pi\sigma,\\
&\Box\epsilon_i=-16\pi\sigma_i,\label{eq27}\\
&\Box h^{\rm TT}_{ij}=-16\pi T^{\rm TT}_{ij}.\label{eq15}
\end{align}

Interestingly the harmonic gauge condition Eqs.~(\ref{eq24}) and (\ref{eq25}) can be expressed with these gauge sensitive variables as
\begin{align}
&\Pi=0,\\
&\frac{2}{3}\partial_i\Lambda-\Sigma_i=0.\label{eq29}
\end{align}
And more within harmonic gauge condition, Eq.~(\ref{eq27}) tells us $\Sigma_i=0$. Consequently Eq.~(\ref{eq29}) becomes
\begin{align}
&\partial_i\Lambda=0.
\end{align}
Because $\Lambda$ goes to zero when $\vec{r}\rightarrow\infty$, the solution of the above equation is
\begin{align}
&\Lambda=0.
\end{align}
In another word, harmonic gauge condition is equivalent to
\begin{align}
&\Pi=0,\label{eq31}\\
&\Lambda=0,\\
&\Sigma_i=0.\label{eq32}
\end{align}

Together with the gauge invariant variables $\Phi$, $\Theta$, $\Xi_i$ and $h^{\rm TT}_{ij}$ determined by Eqs.~(\ref{eq16}), (\ref{eq10}) and (\ref{eq11}), the perturbation metric components can be determined up to initial data of $\epsilon_i$, $\dot{\epsilon}_i$, $\lambda$, $\dot{\lambda}$, $\gamma$ and $\dot{\gamma}$ under harmonic gauge condition. Alternatively we can check Eqs.~(\ref{eq14})-(\ref{eq15}) which can determine the solutions up to initial data. But we need to notice that the initial data for $h^{\rm TT}_{ij}$ correspond to initial gravitational wave content. The initial data for $\epsilon_i$, $\lambda$, and $\gamma$ correspond to the gauge freedom within the harmonic gauge conditions. While the initial data for $\phi$, $H$ and $\beta_i$ should satisfy constrain Eqs.~(\ref{eq31})-(\ref{eq32}) given by the harmonic gauge conditions.

\section{TT gauge condition}\label{sec6}

Eqs.~(\ref{eq14})-(\ref{eq15}) tell us that all the decomposition components $\phi$, $\beta_i$, $\gamma$, $H$, $\lambda$, $\epsilon_i$ and $h^{\rm TT}_{ij}$ are harmonic functions for vacuum case. Eqs.~(\ref{eq10}), (\ref{eq11}) and (\ref{eq16}) tell us that $\Theta=\Xi_i=\Phi=0$ for vacuum case. This fact results in relations
\begin{align}
&H=\nabla^2\lambda,\label{eq17}\\
&\beta_i=\frac{1}{2}\dot{\epsilon}_i,\\
&\phi=\dot{\gamma}-\frac{1}{2}\ddot{\lambda},\label{eq18}
\end{align}
for any harmonic gauge condition if only $T_{\mu\nu}=0$. Starting from any harmonic gauge condition, we can choose
\begin{align}
&C=\frac{1}{2}\lambda,\label{eq33}\\
&B_i=\frac{1}{2}\epsilon_i,\\
&A=\gamma-\frac{1}{2}\dot{\lambda},\label{eq34}
\end{align}
to form $\xi_\mu$ and do a gauge transformation. Since $\lambda$, $\epsilon_i$ and $\gamma$ are harmonic functions, this gauge transformation will result in a harmonic gauge. At the mean time $\lambda$, $\epsilon_i$ and $\gamma$ become zero in the new gauge condition. Eqs.~(\ref{eq17})-(\ref{eq18}) guarantee that $H$, $\beta_i$ and $\phi$ become zero automatically. Consequently $h_{\mu\nu}$ has only $h^{\rm TT}_{ij}$ left. This new gauge condition is nothing but the TT gauge condition.

When matter is presented, Eqs.~(\ref{eq14})-(\ref{eq15}) tell us that all the decomposition components $\phi$, $\beta_i$, $\gamma$, $H$, $\lambda$, $\epsilon_i$ and $h^{\rm TT}_{ij}$ can not be zero, even can not be harmonic functions. So the usual TT gauge condition $h_{\mu\nu}$ equals to $h^{\rm TT}_{ij}$ does not exist in general. But we can still use the gauge transformation Eqs.~(\ref{eq33})-(\ref{eq34}) to set initial data of $\epsilon_i$, $\lambda$, and $\gamma$ to zero. Then the initial data for $\phi$, $H$ and $\beta_i$ are determined by matter through
\begin{align}
&\phi=-\Phi,\\
&H=3\Theta,\\
&\beta_i=\Xi_i.
\end{align}
So when matter is presented, we can call the above initial data choice as generalized TT gauge condition.

\section{CZ gauge condition}\label{sec7}

In \cite{PhysRevD.83.061501,doi:10.1142/S0217732315501928} CZ gauge condition is proposed
\begin{align}
\partial^ih_{i\mu}-\frac{1}{2}\partial_\mu h^i{}_i=0.\label{eq35}
\end{align}
With the decomposition components of perturbation metric, the above condition can be equivalently expressed as
\begin{align}
&\nabla^2\gamma=\frac{1}{2}\dot{H},\label{eq45}\\
&\nabla^2\epsilon_i=\frac{1}{3}\partial_iH-\frac{4}{3}\nabla^2\partial_i\lambda.\label{eq41}
\end{align}
Take divergence of Eq.~(\ref{eq41}) we get
\begin{align}
&0=\frac{1}{3}\nabla^2H-\frac{4}{3}\nabla^2\nabla^2\lambda,\\
&H=4\nabla^2\lambda.\label{eq46}
\end{align}
Consequently Eq.~(\ref{eq41}) becomes
\begin{align}
&\nabla^2\epsilon_i=0,\\
&\epsilon_i=0.\label{eq49}
\end{align}

Under the CZ gauge condition, the linearized Einstein equations reduce to \cite{PhysRevD.83.061501}
\begin{align}
&\nabla^2h_{0\mu}=-16\pi S_{0\mu},\label{eq39}\\
&\nabla^2(\partial_ih^i{}_j)=-16\pi\partial_jT_{00},\label{eq37}\\
&\nabla^2(h^i{}_i)=-32\pi T_{00},\label{eq38}\\
&\Box\hat{h}_{ij}=-16\pi\hat{S}_{ij},\label{eq36}
\end{align}
where $S_{\mu\nu}\equiv T_{\mu\nu}-\frac{1}{2}\eta_{\mu\nu}T$ is the trace-reversed stress-energy tensor with $T$ the trace of $T_{\mu\nu}$. The quantities $\hat{S}_{\mu\nu}$ and $\hat{h}_{\mu\nu}$ are defined by the Poisson equations
\begin{align}
&\nabla^2(S_{\mu\nu}-\hat{S}_{\mu\nu})=\partial_\mu\partial_iS^i{}_{\nu}+\partial_\nu\partial_iS^i{}_{\mu}-\partial_\mu\partial_\nu S^i{}_i,\\
&\nabla^2(h_{\mu\nu}-\hat{h}_{\mu\nu})=\partial_\mu\partial_kh^k{}_{\nu}+\partial_\nu\partial_kh^k{}_{\mu}-\partial_\mu\partial_\nu h^k{}_k.\label{eq42}
\end{align}

Due to Eq.~(\ref{eq35}), we have
\begin{align}
&\partial_\nu\partial^ih_{i\mu}-\frac{1}{2}\partial_\nu\partial_\mu h^i{}_i=0.\label{eq52}
\end{align}
Equivalently we have
\begin{align}
&\partial_\mu\partial^ih_{i\nu}-\frac{1}{2}\partial_\mu\partial_\nu h^i{}_i=0.\label{eq53}
\end{align}
The summation of Eqs.~(\ref{eq52}) and (\ref{eq53}) gives us
\begin{align}
&\partial_\mu\partial^ih_{i\nu}+\partial_\nu\partial^ih_{i\mu}-\partial_\mu\partial_\nu h^i{}_i=0.
\end{align}
Then Eq.~(\ref{eq42}) becomes
\begin{align}
\nabla^2(h_{\mu\nu}-\hat{h}_{\mu\nu})=0,
\end{align}
which means
\begin{align}
\hat{h}_{\mu\nu}=h_{\mu\nu},
\end{align}
under the CZ gauge condition.

Eq.~(\ref{eq39}) can be written as
\begin{align}
&\nabla^2\phi=-4\pi(\rho+3P),\label{eq47}\\
&\nabla^2\beta_i=-16\pi S_i,\label{eq51}\\
&\nabla^2\gamma=-16\pi S,\label{eq50}
\end{align}
which determines $\phi$, $\beta_i$ and $\gamma$ completely.

Eq.~(\ref{eq38}) can be written as
\begin{align}
\nabla^2H=-32\pi\rho,\label{eq48}
\end{align}
which determines $H$ completely. Based on the determined $H$, Eq.~(\ref{eq46}) determined $\lambda$ completely.

Comparing Eqs.~(\ref{eq47})-(\ref{eq48}) and Eqs.~(\ref{eq10})-(\ref{eq16}) we have relations
\begin{align}
&\nabla^2(\phi+\Phi)=\frac{3}{2}\ddot{\Theta},\\
&\beta_i=\Xi_i,\\
&H=4\Theta.
\end{align}
The left $h^{\rm TT}_{ij}$ is controlled by Eq.~(\ref{eq36}). The initial data of $h^{\rm TT}_{ij}$ is determined by the initial gravitational wave content. That is to say the CZ gauge is determined completely by matter content. There is no more gauge freedom in CZ gauge condition.

In vacuum case, Eqs.~(\ref{eq47})-(\ref{eq48}) and (\ref{eq46}) determine $\phi=\beta_i=\gamma=H=\lambda=\epsilon_i=0$ which correspond to the usual TT gauge. So the CZ gauge condition can be viewed as another generalization of the TT gauge condition.

It is interesting to ask if the CZ gauge condition is identical to the generalized TT gauge condition defined in previous section. In order to answer this question we can investigate the behavior of the gauge sensitive variables $\Pi$, $\Lambda$ and $\Sigma_i$ under the CZ gauge condition. Due to Eqs.(\ref{eq45}), (\ref{eq46}) and (\ref{eq49}) we have
\begin{align}
&\Pi=-\dot{\phi},\\
&\Lambda=\frac{3}{2}(\phi-\dot{\gamma}),\\
&\Sigma_i=\dot{\beta}_i.
\end{align}
Eqs.~(\ref{eq47}) and (\ref{eq50}) tell us
\begin{align}
\nabla^2(\phi-\dot{\gamma})=-4\pi(\rho+3P-4\dot{S}).
\end{align}
From the above equation and Eqs.~(\ref{eq47}) and (\ref{eq51}), we can see $\Pi$, $\Lambda$ and $\Sigma_i$ can not be zero in general when matter is presented, which means CZ gauge is harmonic if and only if it is a vacuum. This is to say that the CZ gauge condition is different to the generalized TT gauge condition defined in the previous section when matter is presented.

\section{Conclusion and discussion}

Along with metric perturbation, the TT gauge condition provides a clear physical picture for gravitational waves. It is well known the TT gauge condition is not valid any more when matters present. In the current paper we use scalar-vector-tensor decomposition to analyze harmonic gauge conditions systematically and in depth. Based on our analysis it becomes clear why TT gauge condition can not be satisfied when matters present. Accordingly we generalize the usual TT condition within the harmonic gauge condition framework.

Alternative to harmonic gauge condition, we also analyzed the CZ gauge condition. We find out that CZ gauge condition is different to harmonic in general. The CZ gauge is harmonic when and only when the spacetime is vacuum.

Mathematically, the harmonic gauge conditions are controlled by wave equation while the CZ gauge conditions are controlled by Poisson equation. Since the boundary conditions for the equations are always given by physical conditions, the CZ gauge condition is uniquely determined while the harmonic gauge conditions can be different to each other up to the initial condition for the equation. When the spacetime is vacuum, the uniquely determined CZ gauge condition is nothing but the TT gauge condition. Consequently the CZ gauge condition can be looked as the generalized TT gauge condition when matters are present.

Within TT gauge condition, the perturbation metric components except $h^{\rm TT}_{ij}$ vanishes. This is why people are familiar with the gravitational wave description with $h^{\rm TT}_{ij}$.

When matters are present, the gauge invariant quantity $h^{\rm TT}_{ij}$ is controlled by a wave equation. The mathematical theory tells us $h^{\rm TT}_{ij}\sim\frac{1}{r}$ near null infinity. In contrast the rest gauge invariant quantities $\Phi$, $\Theta$ and $\Xi_i$ are all controlled by Poisson equations. The mathematical theory guarantees that they behaves as $\sim\frac{1}{r^2}$ near null infinity. So near null infinity if we just keep accuracy as $\frac{1}{r}$ order, we can say there is only $h^{\rm TT}_{ij}$ non-zero. Once again it is consistent to our familiar picture that $h^{\rm TT}_{ij}$ describes gravitational waves.

If we use the generalized TT gauge within harmonic gauge conditions when matters are present, all the perturbation metric components are controlled by wave equations. Consequently all of these components behaves as $\sim\frac{1}{r}$ near null infinity.

Differently if we use CZ gauge when matters are present, the perturbation metric component $h^{\rm TT}_{ij}$ is controlled by a wave equation and other components are controlled by Poisson equations. Consequently if we just keep accuracy as $\frac{1}{r}$ order near null infinity, the metric behaves as $h_{\mu\nu}=h^{\rm TT}_{ij}\sim\frac{1}{r}$.
\section*{Acknowledgments}
This work was supported in part by the National Key Research and Development Program of China Grant No. 2021YFC2203001 and in part by the NSFC (No.~11920101003, No.~12021003 and No.~12005016). Z. Cao was supported by ``the Fundamental Research Funds for the Central Universities" of Beijing Normal University. X. He is supported by the NSF of Hunan province (2023JJ30179).

\bibliographystyle{unsrt}
\bibliography{refs}

\begin{thebibliography}{10}

\bibitem{kennefick2007traveling}
Daniel Kennefick.
\newblock {\em Traveling at the Speed of Thought: Einstein and the Quest for
  Gravitational Waves}.
\newblock Princeton university press, 2007.

\bibitem{BonVanMet62}
Hermann Bondi, MGJ Van~der Burg, and AWK Metzner.
\newblock Gravitational waves in general relativity. vii. waves from
  axi-symmetric isolated systems.
\newblock {\em Proceedings of the Royal Society of London. Series A.
  Mathematical and Physical Sciences}, 269(1336):21--52, 1962.

\bibitem{Sac62}
Rainer~K Sachs.
\newblock Gravitational waves in general relativity. viii. waves in
  asymptotically flat space-time.
\newblock {\em Proceedings of the Royal Society of London. Series A.
  Mathematical and Physical Sciences}, 270(1340):103--126, 1962.

\bibitem{2005NJPh....7..204F}
{\'E}anna~{\'E}. {Flanagan} and Scott~A. {Hughes}.
\newblock {The basics of gravitational wave theory}.
\newblock {\em New Journal of Physics}, 7(1):204, September 2005.

\bibitem{2022GReGr..54...89C}
Rong-Gen {Cai}, Xing-Yu {Yang}, and Long {Zhao}.
\newblock {On the energy of gravitational waves}.
\newblock {\em General Relativity and Gravitation}, 54(8):89, August 2022.

\bibitem{liangcanbin}
Canbin Liang and Bin Zhou.
\newblock {\em Introductory differential geometry and general relativity I, II,
  III}.
\newblock Science Press Pub press, 2000.

\bibitem{wald84}
Robert~M. Wald.
\newblock {\em General relativity}.
\newblock The University of Chicago Press, Chicago, 1984.

\bibitem{MTW73}
Charles~W. Misner, Kip~S. Thorne, and John~Archibald Wheeler.
\newblock {\em Gravitation}.
\newblock Princeton University Press, 1973.

\bibitem{PhysRevD.94.024048}
Fan Zhang.
\newblock Accumulative coupling between magnetized tenuous plasma and
  gravitational waves.
\newblock {\em Phys. Rev. D}, 94:024048, Jul 2016.

\bibitem{2022CQGra..39g5014L}
Richard {Lieu}, Kristen {Lackeos}, and Bing {Zhang}.
\newblock {Damping of long wavelength gravitational waves by the intergalactic
  medium}.
\newblock {\em Classical and Quantum Gravity}, 39(7):075014, April 2022.

\bibitem{maggiore2008gravitational}
Michele Maggiore.
\newblock {\em Gravitational waves: Volume 1: Theory and experiments}.
\newblock Oxford university press, 2008.

\bibitem{Li:2023plm}
Junlang Li, Fangfei Liu, Yuan Pan, Zijian Wang, Mengdi Cao, Mengyao Wang, Fan
  Zhang, Jinhai Zhang, and Zong-Hong Zhu.
\newblock {Detecting gravitational wave with an interferometric seismometer
  array on lunar nearside}.
\newblock {\em Sci. China Phys. Mech. Astron.}, 66(10):109513, 2023.
\newblock [Erratum: Sci.China Phys.Mech.Astron. 67, 219551 (2024)].

\bibitem{PhysRevD.110.043009}
Han Yan, Xian Chen, Jinhai Zhang, Fan Zhang, Lijing Shao, and Mengyao Wang.
\newblock Constraining the stochastic gravitational wave background using the
  future lunar seismometers.
\newblock {\em Phys. Rev. D}, 110:043009, Aug 2024.

\bibitem{2009arXiv0912.0128R}
Istvan {Racz}.
\newblock {Gravitational radiation and isotropic change of the spatial
  geometry}.
\newblock {\em arXiv e-prints}, page arXiv:0912.0128, December 2009.

\bibitem{PhysRevD.83.061501}
Xiang-Song Chen and Ben-Chao Zhu.
\newblock True radiation gauge for gravity.
\newblock {\em Phys. Rev. D}, 83:061501, Mar 2011.

\bibitem{doi:10.1142/S0217732315501928}
Ben-Chao Zhu and Xiang-Song Chen.
\newblock Tensor gauge condition and tensor field decomposition.
\newblock {\em Modern Physics Letters A}, 30(35):1550192, 2015.

\end{thebibliography}

\end{document}